\documentclass{llncs}

\usepackage[dvips]{epsfig}
\usepackage{epsfig}
\usepackage{graphicx}
\usepackage{color}
\usepackage{fullpage}
\usepackage{latexsym}
\usepackage{amsmath}
\usepackage{amssymb}

\long\def\commabs #1\commabsend{}

\newcommand{\reals}{\hbox{$\rlap{\rm I} \> \kern-.2mm{\rm R}$}}

\newcommand{\VLCOVER}{\mbox{\bf VL\_Cover}}
\newcommand{\WVLCOVER}{\mbox{\bf WVL\_Cover}}
\commabs

\newcommand{\VLCOVER}{\mbox{\bf VL\_Cover}}
\commabsend

\newcommand{\dist}{\mbox{\bf dist}}

\def\inline#1:{\par\vskip 7pt\noindent{\bf #1:}\hskip 10pt}
\def\Proof{\par\noindent{\bf Proof:~}}

\newcommand{\E}{\mathop{\mathbb E}}

\def\inline#1:{\par\vskip 7pt\noindent{\bf #1:}\hskip 10pt}
\def\Proof{\par\noindent{\bf Proof:~}}
\def\blackslug{\hbox{\hskip 1pt \vrule width 4pt height 8pt
    depth 1.5pt \hskip 1pt}}

\def\QED{\quad\blackslug\lower 8.5pt\null\par}

\begin{document}
\title{Improved Distance Oracles and Spanners for Vertex-Labeled Graphs}

\author{
Shiri Chechik\inst{1}}

\institute{
Department of Computer Science, The Weizmann Institute, Rehovot, Israel\\
\email{shiri.chechik@weizmann.ac.il}
}

%

%

\maketitle

\begin{abstract}
Consider an undirected weighted graph $G=(V,E)$ with $|V| = n$ and $|E| = m$,
where each vertex $v \in V$ is assigned a {\em label} from a set of labels $L = \{\lambda_1,...,\lambda_{\ell}\}$.
We show how to construct a compact distance oracle that can answer queries of the form:
``what is the distance from $v$ to the closest $\lambda$-labeled node" for a given node $v\in V$ and label $\lambda \in L$.

This problem was introduced by Hermelin, Levy, Weimann and Yuster [ICALP 2011] where they present several results for this problem.
In the first result, they show how to construct a vertex-label distance oracle of expected size $O(kn^{1+1/k})$ with stretch $(4k - 5)$ and query time $O(k)$.
In a second result, they show how to reduce the size of the data structure to $O(kn \ell^{1/k})$ at the expense of a huge stretch,
the stretch of this construction grows exponentially in $k$, $(2^k-1)$.
In the third result they present a dynamic vertex-label distance oracle that is capable of handling label changes in a sub-linear time. The stretch of this construction is also exponential in $k$, $(2\cdot 3^{k-1}+1)$.

We manage to significantly improve the stretch of their constructions, reducing the
dependence on $k$ from exponential to polynomial $(4k-5)$, without requiring any tradeoff regarding any of the other variables.

In addition, we introduce the notion of vertex-label spanners: subgraphs that preserve distances between every node $v \in V$ and label $\lambda \in L$. We present an efficient construction for vertex-label spanners with stretch-size tradeoff close to optimal.
\end{abstract}


\section{Introduction}

An {approximate distance oracle} for a given graph $G = (V,E)$ is a processed
data structure that, given two nodes $s$ and $t$, can quickly return an approximation of $\dist(s,t,G)$, the distance between $s$ and $t$ in $G$.
[To ease notation, we let $\dist(s,t) = \dist(s,t,G)$. In other words, when we refer to the distance between $s$ and $t$ in some subgraph $H$ of $G$, we will always state the subgraph explicitly and write $\dist(s,t,H)$. Otherwise, if we write $\dist(s,t)$ we mean $\dist(s,t,G)$].

The approximate distance oracle is said to be of stretch $k$, or a $k$-approximate distance oracle, if for every two nodes, $s$ and $t$, the reported distance $\tilde{\dist}(s,t)$ between $s$ and $t$ satisfies
$\dist(s,t) \leq  \tilde{\dist}(s,t) \leq k \cdot \dist(s,t)$.

Usually, the key concerns in designing approximate distance oracles are to minimize the size of the data structure, to minimize the stretch, and to minimize the query time.

Distance oracles have been extensively studied.
They were first introduced by Thorup and Zwick in a seminal paper \cite{ThZw05}.
Thorup and Zwick showed how to construct for a given integer $k \geq 1$, a $(2k-1)$-approximate distance oracle of size $O(kn^{1+1/k})$ that can answer distance queries in $O(k)$ time.
Thorup and Zwick \cite{ThZw05} showed that their space requirements are essentially optimal assuming the girth conjecture of Erd\H{o}s \cite{Er64}.
Thorup and Zwick also showed how to derandomize this construction, but at the cost of increasing the preprocessing time and slightly increasing the size. Roditty, Thorup, and Zwick \cite{RoThZw05} later improved this result, presenting
a faster deterministic construction and reducing the size of the data structure to
$O(kn^{1+1/k})$ as in the randomized construction.
Further improvements on the construction time were later introduced in \cite{BaKa06,BaSe06,BaGaSeUp08}.
For further results and lower bounds see also \cite{MeNa06,SoVeYu09,PaRo10,AG11}.

In this paper, we consider a natural variant of the approximate distance oracle problem for vertex-labeled graphs.
We are given an undirected weighted graph, $G=(V,E)$, where each vertex, $v$, is assigned a {\em label}, $\lambda(v)$, where $\lambda(v)$ belongs to a set $L = \{\lambda_1,...,\lambda_{\ell}\}$ of $\ell \leq n$ distinct labels.
The goal is to construct a compact data structure that, given a node $v$ and a label $\lambda \in L$, can quickly return an approximation to the distance $\dist(v,\lambda)$, where $\dist(v,\lambda)$ is the minimal distance between $v$ and a $\lambda$-labeled node in $G$.
This interesting variant of distance oracles was introduced by Hermelin, Levy, Weimann and Yuster \cite{HeLeWeYu11}.
The labels of the nodes often represent some functionality (or resources).
In some settings, the natural question is not what is the distance between two given nodes, but
rather what is the distance between a given node and some desired resource.
For example, the nodes may represent cities and the labels may represent some public resources such as hospitals, courts, universities, and so on.

Hermelin et al. \cite{HeLeWeYu11} mention that there is a simple solution for this problem: store a table of size $n \cdot \ell$, where the entry $(v,\lambda)$ represents the distance $\dist(v,\lambda)$.
This data structure is of size $O(n \ell)$, the query time is $O(1)$, and the stretch is 1 (exact distances). As shown in \cite{HeLeWeYu11} this table can be constructed in $O(m \ell)$ time. However, as is also mentioned in \cite{HeLeWeYu11}, this data structure suffers from two main drawbacks.
First, in many applications, $O(n \ell)$ might be still too large,
and in such applications it might be  preferable to store a more compact data structure at the price of approximate distances.
Second, in some settings it might be desirable to allow label changes and it is not clear how to efficiently handle label changes using the above mentioned data structure.

Hermelin et al. \cite{HeLeWeYu11} present several results for distance oracles for vertex-labeled graphs problem.
In their first result, Hermelin et al. show how to construct a vertex-label distance oracle of expected size $O(kn^{1+1/k})$ with stretch $(4k - 5)$ and query time $O(k)$.
This result is unsatisfactory when $\ell$ is very small, especially when $\ell = o(n^{1/k})$. In this case, the trivial $O(n \ell)$ solution gives a smaller data structure with exact distances.
To overcome this issue, they propose a second data structure of size $O(kn \ell^{1/k})$. This, however, comes at the price of a huge stretch factor of $2^k-1$.
In a third result, they present a dynamic vertex-label distance oracle that is capable of handling label changes in sub-linear time.
More specifically, they show how to construct a vertex-label
distance oracle of expected size $O(kn^{1+1/k})$ and with stretch ($2\cdot 3^{k-1}+1)$ that can support label changes in $O(kn^{1/k} \log\log n)$ time and queries in $O(k)$ time.

Note that in the latter two results, the stretch depends exponentially on $k$.
In this paper, we address an important question they left open, namely, is it possible to improve this dependence on $k$ from exponential to polynomial.
More specifically, we prove the following theorems.


\begin{theorem}
\label{the:staticOracle}
A vertex-label distance oracle of expected size $O(k n \ell^{1/k})$ with stretch $(4k-5)$ and query time $O(k)$ can be constructed in
$O(m \cdot \min\{n^{k/(2k-1)},\ell\})$ time.
\end{theorem}

\begin{theorem}
\label{the:dynamicOracle}
A vertex-label distance oracle of expected size $\tilde{O}(n^{1+1/k})$ with stretch $(4k-5)$ and query time $O(k)$ can
be constructed in $O(k m n^{1/k})$ time and can support label changes in $O(n^{1/k} \log^{1-1/k}{n} \log\log{n})$ time.
\end{theorem}

A closely related notion of distance oracles is that of \emph{spanners}.
A subgraph $H$ is said to be a $k$-spanner (or a spanner with stretch $k$) of the graph $G$ if $\dist(u,v,H) \leq k \cdot \dist(u,v,G)$ for every $u,v \in V(G)$.
Here and throughout, $V(G')$ denotes the set of vertices of graph $G'$, and similarly, $E(G')$ denotes the set of edges of graph $G'$.
A well-known theorem on spanners is that one can efficiently construct a $(2k-1)$-spanner
with $O(n^{1+1/k})$ edges \cite{ADDJS93}.
This size-stretch tradeoff is conjectured to be optimal .
The notion of spanners was introduced in the late 80's \cite{PelegS-89,PU89b},
and has been extensively studied.
Spanners are used as a fundamental ingredient in many distributed applications (e.g., synchronizers~\cite{PU89b},
compact routing~\cite{PU89,ThZw01}, broadcasting~\cite{FPZW04}, etc.).

This paper also introduces a natural extension of spanners, spanners for vertex-labeled graphs, and presents efficient constructions for such spanners.
Consider an undirected weighted graph $G=(V,E)$, where each vertex $v \in V$ is assigned a label from a set of labels $L = \{\lambda_1,...,\lambda_{\ell}\}$.
We say that a subgraph $H$ is a {\em vertex-labeled $k$-spanner (VL $k$-spanner)} of $G$ if $\dist(u,\lambda,H) \leq k \cdot \dist(u,\lambda,G)$ for every node $u \in V$ and label $\lambda \in L$.
It is not hard to verify that every $k$-spanner is also a VL $k$-spanner. However, one may hope to find sparser $VL$-spanners when the number of labels is small.
A naive approach would be to create for each label $\lambda$ an auxiliary graph $G_{\lambda}$ by adding a new node $s_{\lambda}$ and then connect
$s_{\lambda}$ to all $\lambda$-labeled nodes with edges of weight 0.
It is not hard to verify that by invoking a shortest path algorithm in every $G_{\lambda}$ from $s_{\lambda}$
and taking the union of all these shortest-paths trees (removing the nodes $s_{\lambda}$ and their incident edges), the resulting subgraph is a VL 1-spanner (preserving the exact distances) with $O(n \ell)$ edges.
However, the $O(n \ell)$ spanner's size may still be too large in many settings , and one may wish to reduce the size of the spanner at the price of
approximated distances. Ideally, one would wish to find a VL $(2k-1)$-spanner with $O(n \ell^{1/k})$ number of edges (beating these bounds yields improved trade-off for the standard spanners).
We managed to come close this goal, presenting an efficient construction for VL spanners with stretch close to $(4k+1)$ and with $\tilde{O}(n \ell^{1/k})$ number of edges.
More specifically, we prove the following theorem.

\begin{theorem}
\label{the:weight-spanner}
For every weighted graph $G$ with minimal edge weight 1 and fixed parameter $\epsilon > 0$, one can efficiently construct a vertex-label $(4k+1)(1+\epsilon)$-spanner
with $O(\log{n} \cdot \log{D} \cdot n \ell^{1/k})$ edges, where $D$ is the diameter of the graph.
\end{theorem}

We note that our constructions for vertex-labeled distance oracles and the constructions presented in \cite{HeLeWeYu11} do not seem to transform well to also give spanners.
Therefore, our vertex-labeled spanner constructions use different techniques and require some new ideas (and this is the technically more involved part of this paper, Section~\ref{sec:spanners}).

The rest of the paper is organized as follows.
In Section~\ref{sec:compact-oracle} we prove Theorem \ref{the:staticOracle}.
In Section~\ref{sec:dynamic-oracle} we prove Theorem \ref{the:dynamicOracle}.
In Section~\ref{sec:spanners} we prove Theorem \ref{the:weight-spanner}, for simplicity, we first present a construction for unweighted graphs in
Subsection \ref{sec:u-spanners} and then show how to generalize it to weighted graphs in Subsection \ref{sec:w-spanners}.

\section{Compact Vertex-Label Distance Oracles}
\label{sec:compact-oracle}
In this section we prove Theorem \ref{the:staticOracle}.
In Subsection \ref{subsec:data-strcuture-static} we present the construction of our data structure, in  Subsection \ref{subsec:query-static} we present our query answering algorithm, and in
Subsection \ref{subsec:const-static} we analyze the construction time.

\subsection{The Data Structure}
\label{subsec:data-strcuture-static}
The first step of the construction of the data structure is similar to the algorithm presented in \cite{HeLeWeYu11}.
For a given positive integer $k$, construct the sets $V = A_0 \supseteq A_1 \supseteq \cdots \supseteq A_{k-1}$ as follows:
The $i$-th level $A_i$ is constructed by sampling the vertices of $A_{i-1}$ independently at random with probability $\ell^{-1/k}$ for $1\leq i\leq k-1$.

Next, for every vertex $v$, define the bunch of $v$ exactly as the Thorup-Zwick definition, but with a small change: that is, omit the last level, namely
$$B(v) = \bigcup_{i=0}^{k-2}{\{u \in A_i \setminus A_{i+1} \mid \dist(v,u) < \dist(v,A_{i+1})\}}.$$
The pivot $p_i(v)$ is also exactly as Thorup-Zwick's definition, namely $p_i(v)$ is the closest node to $v$ in $A_i$ (break ties arbitrarily).


Next, for every node $v \in A_{k-1}$, store its distance for every label $\lambda \in L$, namely $\dist(v,\lambda)$ in a hash table.
Finally, for every label $\lambda \in L$, store $B(\lambda) = \bigcup_{v \in V_{\lambda}}{B(v)}$ in a hash-table and for every node $x \in B(\lambda)$ store $\dist(x,\lambda)$.


This completes the description of our data structure.

Below, we bound the size of the data structure.

\begin{lemma}
\label{lem:ball-size}
$\E[B(v)] = (k-1) \ell^{1/k}$.
\end{lemma}
\Proof
Using the same analysis as Thorup-Zwick's, one can show that
the expected size of $B(v) \cap (A_i\setminus A_{i+1})$, for $1 \leq i \leq k-2$,
is stochastically dominated by a geometric random variable with
parameter $p = \ell^{-1/k}$.


Hence, $\E[B(v)  \cap A_i\setminus A_{i+1}] =  \ell^{1/k}$.
We thus get that $\E[B(v)] = (k-1) \ell^{1/k}$.
\QED

\begin{lemma}
The expected size of our data structure is $O(k n \ell^{1/k})$.
\end{lemma}
\Proof
By Lemma \ref{lem:ball-size}, the total expected size of the bunches of all nodes is $(k-1) n \ell^{1/k}$.
In addition, for every node in $A_{k-1}$ we also store its distance to every label, $\lambda \in L$. That is, for every node in $A_{k-1}$, we store additional data of size $\ell$.
The expected size of $A_{k-1}$ is $n \ell^{-(k-1)/k}$. To see this, note that the probability that a node $v$ belongs to $A_i$ is $\ell^{-i/k}$.
Therefore, the total additional expected size stored for all nodes in $A_{k-1}$ is $n \ell^{1/k}$.
Finally, storing $B(\lambda)$ for every $\lambda \in L$ does not change the asymptotic size since $\sum\limits_{\lambda\in L}{|B(\lambda)|} = \sum\limits_{\lambda\in L}{\sum\limits_{v \in V_{\lambda}}{|B(v)|}} = \sum\limits_{v \in V}{|B(v)|}$.
The lemma follows.
\QED

\subsection{Vertex-Label Queries}
\label{subsec:query-static}
We now describe our query answering algorithm, with the input vertex-label query, $(v \in V, \lambda \in L)$.

The query answering algorithm is done as follows.
For every index $i$ from $0$ to $k-2$, check if $p_i(v) \in B(\lambda)$, and if so return $\dist(v,p_i(v)) + \dist(p_i(v),\lambda)$.
Otherwise, if no such index exists, return $\dist(v,p_{k-1}(v)) + \dist(p_{k-1}(v),\lambda)$.
This completes the query answering algorithm.

We now turn to the stretch analysis.
If there exists an index $i$ such that $0 \leq i \leq k-2$
and $p_i(v) \in B(\lambda)$, set $i$ to be the first such index. If no such index exists, set $i = k-1$.
Let $u$ be the $\lambda$-labeled node closest to $v$, namely $\dist(v,u) = \dist(v,\lambda)$.
Note that $p_j(v) \notin B(u)$ for every $j < i$. This is due to the facts that
$p_j(v) \notin B(\lambda)$ and that $B(u) \subseteq B(\lambda)$.
Using the same analysis as in \cite{ThZw01} (Lemma $A.1$), one can show that $\dist(v,p_{i}(v)) \leq (2i-2)\dist(v,u)$ and
$\dist(p_i(v),\lambda) \leq \dist(p_i(v),u) \leq (2i-1))\dist(v,u)$.
We get that the returned distance $\tilde{\dist}(v,\lambda)$ satisfies
$\tilde{\dist}(v,\lambda) = \dist(v,p_{i}(v)) + \dist(p_i(v),\lambda) \leq (4k-3)\dist(v,u) = (4k-3)\dist(v,\lambda)$.

Note that if $i \leq k-2$, then the distance $\dist(p_i(v),\lambda)$ is stored in $B(\lambda)$, or,
if $i = k-1$ then $p_i(v) \in A_{k-1}$ and
recall that $\dist(u,\lambda)$ is stored for every $u \in A_{k-1}$ and therefore also $\dist(p_i(v),\lambda)$ is stored and can be retrieved in $O(1)$ time.

Finally, using the same method as in \cite{ThZw01} (Lemma $A.2$) the stretch can be reduced to $4k-5$ as required.

We note that Hermelin et al. \cite{HeLeWeYu11} have to check all indices.
Namely, their query algorithm is to return the minimal distance
$\dist(v,w) + \dist(w,w_\lambda)$ for all $w = p_i(v)$ such that $w \in B(\lambda)$, where we define $w_\lambda$ to be the $\lambda$-labeled node closest to $w$ that satisfies $w\in B(w_\lambda)$.
Let $u$ be the $\lambda$-labeled node that satisfies $\dist(v,u) = \dist(v,\lambda)$.
Hermelin et al. \cite{HeLeWeYu11} note that the first $w =p_i(v) \in B(\lambda)$, does not necessarily satisfy $\dist(w,w_\lambda) \leq \dist(w,u)$ since there is a possibility that $w \notin B(u)$.
Therefore, they have to iterate over all indices $1 \leq i \leq k-1$ and take the one that gives the minimal distance.
We bypass this issue by simply explicitly storing the distance $\dist(w,\lambda)$ ---rather than $\dist(w,w_\lambda)$--- for every $w \in B(\lambda)$.
This does not increase the asymptotic size and it simplifies the query algorithm and its analysis.

\subsection{Construction Time}
\label{subsec:const-static}
The preprocessing time of our construction is composed of the time it takes to construct
the different components of our data structure.
Recall that our data structure is composed of four components.
The first component is the pivots: for every node $v$ we store $p_i(v)$ for $1 \leq i \leq k-1$.
The second component is the bunches of the vertices: for every node $v$ we store $B(v)$ and the distances $\dist(v,x)$ for every $x \in B(v)$.
The third component is the bunches of the labels: for every label $\lambda$ we store $B(\lambda)$ and the distances $\dist(x,\lambda)$ for every $x \in B(\lambda)$.
The fourth part is the distances of the nodes in $A_{k-1}$ to all labels: store $\dist(v,\lambda)$ for every $v \in A_{k-1}$ and $\lambda \in L$.

Using the same analysis as in \cite{ThZw05}, one can show that the time complexity for constructing the first component is $O(k \cdot m)$ and the time complexity for the second component is $O(k m \ell^{1/k})$.

Constructing $B(\lambda)$ for every $\lambda \in L$ (the first part of the third component) can be done easily in $O(k n \ell^{1/k})$ time [just go over all nodes $v$, and add $B(v)$ to $B(\lambda(v))$].

We are left with computing $\dist(x,\lambda)$ for every $x \in B(\lambda)$ and then for every $x \in A_{k-1}$ and $\lambda \in L$.
This can be done by invoking Dijkstra's Algorithm $\ell$ times (for every label $\lambda \in L$ add a source node $s$ and connect all $\lambda$-labeled nodes to $s$ with an edge of weight 0 and then invoke Dijkstra's Algorithm from $s$) and thus the running time for this part is $O(m \ell)$.

We get that the total running time for the preprocessing phase is $O(m\ell)$.

We note here that if $\ell > n^{k/(2k-1)}$, then it is possible to reduce the preprocessing running time to $O(m n^{k/(2k-1)})$.
This can be done by storing $\dist(v,v_\lambda)$ as suggested in \cite{HeLeWeYu11} instead of storing $\dist(v,\lambda)$ for every $v \in B(\lambda)$. This change forces checking all indices in the query algorithm as explained above.
The analysis of the preprocessing time in this case is similar to the one presented in \cite{HeLeWeYu11}.

%
%
%
%
%
%

\section{Dynamic Labels}
\label{sec:dynamic-oracle}
In this section, we consider the problem of constructing a dynamic vertex-label distance oracle and prove Theorem \ref{the:dynamicOracle}.
Namely, we show how to construct a vertex-label distance oracle that supports label changes of the form $update(v,\lambda)$ for $v\in V$ and $\lambda \in L$.
This update changes the label of $v$ to be $\lambda$ and leaves all other nodes unchanged.
Our data structure in this section is a slight adaptation of Thorup-Zwick's construction and is also similar to the one presented in \cite{HeLeWeYu11} for static vertex-label distance oracles.

In Subsection \ref{subsec:data-strcuture-dynamic} we present the data structure, Subsection \ref{subsec:query-dynamic} presents the query answering algorithm and in
Subsection \ref{subsec:const-dynamic} we analyze the construction time.

\subsection{The Data Structure}
\label{subsec:data-strcuture-dynamic}
For a given positive integer $k$, construct the sets $V = A_0 \supseteq A_1 \supseteq \cdots \supseteq A_{k-1} \supseteq A_k = \emptyset$ as follows.
The $i$-th level $A_i$ is constructed by sampling the vertices of $A_{i-1}$ independently at random with probability $p$ to be specified shortly for $1\leq i\leq k-1$.

The bunch of $v$ is defined as in Thorup-Zwick as follows:
$$B(v) = \bigcup_{i=0}^{k-1}{\{u \in A_i \setminus A_{i+1} \mid \dist(v,u) < \dist(v,A_{i+1})\}}.$$
The pivot $p_i(v)$ is also defined exactly as Thorup-Zwick's definition, namely $p_i(v)$ is the closest node to $v$ in $A_i$ (break ties arbitrarily).

In order to allow fast updates, the size of every bunch $B(v)$ for $v \in V$ must be small.
In order to ensure this property, we set the sampling probability to be $p = (n/\ln{n})^{-1/k}$.
It was proven in \cite{ThZw05} that by setting $p = (n/\ln{n})^{-1/k}$, the size of every bunch
$B(v)$ is $O(n^{1/k} \log^{1-1/k}{n})$ with high probability .

In addition, for every $\lambda \in L$,
store $B(\lambda) = \bigcup_{v \in V_{\lambda}}{B(v)}$ in a hash-table.
Recall that in the static setting we store $\dist(v,\lambda)$ when $v \in B(\lambda)$.
In the dynamic setting, we do not store this data as it is too costly to update it for two reasons.
First, notice that a single label change, say from $\lambda_1$ to $\lambda_2$, might require updating $\dist(v,\lambda_1)$, $\dist(u,\lambda_2)$
for many nodes $v \in B(\lambda_1)$ and $u \in B(\lambda_2)$.
As both $B(\lambda_1)$ and $B(\lambda_2)$ might be very large, this may take a long time.
Second, even a single update of $\dist(v,\lambda)$ might be too costly as it might require
invoking a shortest path algorithm during the update phase.

To avoid the need of updating $\dist(v,\lambda)$ for a node $v \in B(\lambda)$, we do the following two things.
First, rather than maintaining the value $\dist(v,\lambda)$, we instead maintain the value
$\dist(v,v_\lambda)$ where $v_\lambda$ is defined to be the closest $\lambda$-labeled node such that $v \in B(v_\lambda)$.
Second, we use the method of \cite{HeLeWeYu11} and iterate on all indices $1\leq i \leq k-1$ and return the minimal distance
$\dist(v,w) + \dist(w,w_\lambda)$ for $w = p_i(v)$ in the answering query algorithm.

In order to maintain the value $\dist(v,w_\lambda)$ for a node $v \in B(\lambda)$,
we store the set of $\lambda$-labeled nodes $x$ such that $v$ belongs to $B(x)$ in a heap, $Heap(v,\lambda)$, namely
the set of nodes in the $Heap(v,\lambda)$ is $V(Heap(v,\lambda)) = \{x\in V \mid v \in B(x) ~and~ \lambda(x) = \lambda\}$ where the key, $key(x)$, of a node, $x \in V(Heap(v,\lambda))$, is the distance, $\dist(v,x)$.
The heap, $Heap(v,\lambda)$, supports the standard operations of $[insert(x)$ - insert a node $x$ to the heap], [$remove(x)$ - remove a node $x$ from the heap] and [$minimum()$ -
return the node $x$ in the heap with minimal $key(x)$].
For this purpose, we use any standard construction of heaps (e.g. \cite{VaKaZi77}) that allow $insert$ and $remove$ operations in $O(\log\log{n})$ time and $minimum()$ operations at constant time.

We now summarize the different components of our data structure to make it clear what parts of the data structure need to be updated as a result of a label change.

\begin{description}
\item{(1)} For every node $v$, store $B(v)$ and for every node $x \in B(v)$, store $\dist(v,x)$.
This data is stored in a hash-table, which allows checking if a node $x \in B(v)$ and, if so, finding $\dist(v,x)$ in $O(1)$ time.
\item{(2)} For every node $v$ and index $1\leq i\leq k-1$, store $p_i(v)$.
\item{(3)} For every $\lambda \in L$, store $B(\lambda)$ in a hash-table where the entry in the hash-table of a node $v \in B(\lambda)$ points to the heap, $Heap(v,\lambda)$.
\end{description}

It is not hard to see that only component $(3)$ in our data structure needs to be modified as a result of a label change.
Moreover, if the label of some node $v \in V$ is changed from $\lambda_1 \in L$ to $\lambda_2 \in L$, then only $B(\lambda_1)$ and $B(\lambda_2)$ need to be updated.
The update is relatively simple.
For every node $x \in B(v)$, do the following:
Remove $v$ from $Heap(x,\lambda_1)$. If $Heap(x,\lambda_1)$ becomes empty, then also remove $x$ from the hash-table of $B(\lambda_1)$.
In addition, if $x \in B(\lambda_2)$, then add $v$ to $Heap(x,\lambda_2)$, otherwise add $x$ to the hash-table $B(\lambda_2)$ and create a new heap,
$Heap(x,\lambda_2)$, containing only $v$.
Each such operation takes $O(\log\log{n})$ time for every $x \in B(v)$.
Recall that the size of $B(v)$ is $O(n^{1/k} \log^{1-1/k}{n})$; thus we get that the update requires $O(n^{1/k} \log^{1-1/k}{n} \log\log{n})$ time.

It is not hard to verify that the size of the data structure is $O(\sum\limits_{v\in V}{|B(v)|}) = O(n^{1+1/k} \log^{1-1/k}{n})$.

\subsection{Vertex-Label Queries}
\label{subsec:query-dynamic}
The query answering algorithm is similar to the one presented in Section \ref{sec:compact-oracle}.
Let $(v \in V, \lambda \in L)$ be the input vertex-label query.

The query answering algorithm is done by checking all indices $1\leq i \leq k-1$ and
returning the minimal $\dist(v,p_i(v)) + \dist(p_i(v), w_\lambda)$ such that
$p_i(v) \in B(\lambda)$. $w_\lambda$ is the node returned by $Heap(p_i(v),\lambda).minimum()$, namely,
$w_\lambda$ is the $\lambda$-labeled node such that $p_i(v) \in B(\lambda)$ with minimal $\dist(p_i(v), w_\lambda)$.

Note that here we must check all indices and cannot stop upon reaching the first index $j$, such that $p_j(v) \in B(\lambda)$. Let $u$ be the $\lambda$-labeled node closest to $v$, namely $\dist(v,u) = \dist(v,\lambda)$.
As mentioned by Hermelin et al. \cite{HeLeWeYu11} (and discussed above), the first $w =p_i(v) \in B(\lambda)$, does not necessarily satisfy $\dist(w,w_\lambda) \leq \dist(w,u)$ since it may be that $w \notin B(u)$.
Therefore, we also have to iterate over all indices $1 \leq i \leq k-1$ and take the one that gives the minimal distance.

It is not hard to verify that the query algorithm takes $O(k)$ time.

Finally, as mentioned in Section \ref{sec:compact-oracle}, using the same analysis as in \cite{ThZw01} (Lemma $A.1$),
the stretch is $(4k-3)$, and using the same method as in \cite{ThZw01} (Lemma $A.2$), the stretch can be reduced to $4k-5$ as required.

\subsection{Construction Time}
\label{subsec:const-dynamic}
The first two components of the data structure are exactly the same construction as Thorup-Zwick's and thus can be constructed in $O(k m n^{1/k})$ time.
The third component can be constructed in $O(|\cup_{v \in V}{B(v)| \cdot \log\log{n}}) = O(k n^{1+1/k} \log\log{n})$ time (the $\log\log{n}$ comes from the insertion to the heaps).
We conclude that the total preprocessing time is $O(k m n^{1/k})$.

\section{Sparse Vertex-Label Spanners}
\label{sec:spanners}
In this section, we shall address the question of finding low stretch sparse vertex-label spanners.
More specifically, we show how to find a subgraph $H$ with expected number of edges $\tilde{O}(n \ell^{1/k})$
such that for every vertex $v$ and label $\lambda$, $\dist(v,\lambda,H) \leq (4k+1) (1+\epsilon)\dist(v,\lambda,G)$ for any fixed $0 < \epsilon$.
Note that it is unclear how to transform the construction of Section \ref{sec:compact-oracle} into a vertex-label spanner.
To see this, recall that for every node $v$ in $A_{k-1}$ and for every label $\lambda$ we store the distance $\dist(v,\lambda)$.
However, in order to allow a low-stretch spanner, we need to add a shortest path $P$ from $v \in A_{k-1}$ to its closest $\lambda$-labeled node. This path could be very long and, of course, may contain many nodes not in $A_{k-1}$.
Thus, adding all these paths may result with a subgraph with too many edges.
Therefore, transforming the construction of Section \ref{sec:compact-oracle} into a vertex-label spanner seems challenging.
We hence suggest a different construction for spanners.

For simplicity, we first present (Subsection \ref{sec:u-spanners}) a construction for unweighted graphs and then (Subsection \ref{sec:w-spanners}) we show how to generalize this construction to weighted graphs.

\subsection{Unweighted Graphs}
\label{sec:u-spanners}

For a node $v$, radius $r$, and subgraph $H$, let $B(v,r,H) = \{x \in V(H) \mid \dist(v,x,H) \leq r\}$.

We start by describing an algorithm named \VLCOVER, that when given a distance $d$, it returns a spanner $H$ with the following property.
For every node $v \in V$ and label $\lambda \in L$, such that $\dist(v,\lambda,G) \leq d$, $\dist(v,\lambda,H) \leq (4k+1)d$.

Loosely speaking, the algorithm proceeds as follows:
It consists of two stages.
The first stage handles ``sparse'' areas, namely, balls around some node $v$ such that $|B(v, k d, G)| < d \cdot \ell^{1-1/k}$.
In this case, we show that we can tolerate adding a BFS tree $T(v)$ spanning $B(v, i d, G)$ for some $1 \leq i \leq k$, charging the nodes in
$B(v, (i-1) d, G)$. It is not hard to verify that every path $P$ of length at most $d$ that contains a node in $B(v, (i-1) d, G)$ satisfies $V(P) \subseteq B(v, i d, G)$. Using the tree $T(v)$, we have a ``short'' alternative path to $P$.
The second stage handles ``dense'' areas, namely, balls around some node $v$ such that $|B(v, k d, G)| \geq d \cdot \ell^{1-1/k}$.
The algorithm picks a set $C \subseteq V$ such that
the distance between every two nodes in $C$ is at least $2 k d$, and that every node in a  ``dense'' area has a  ``close'' node in $C$.
In this case, we can tolerate adding $O(d \cdot \ell)$ edges for every node $c\in C$, charging the nodes in $B(c, k d, G)$.
The algorithm connects every node $u \in V$ to some ``close'' node $c \in C$ by a shortest path.
In addition, for every label $\lambda$ such that the distance from $c$ to $\lambda$ is $O(d)$, we add a ``short'' path from $c$ to $\lambda$.
In this case for every node $u$ and label $\lambda$ such that $\dist(u,\lambda) \leq d$, we have  a ``short'' alternative path by concatenating the path from $u$ to its ``close'' node $c \in C$ and the path from $c$ to $\lambda$.

We now describe the algorithm more formally.
\par
Initially, set $G' =G$, $H_d =(V,\emptyset)$, and $C = \emptyset$.
The algorithm consists of two stages.
The first stage of the algorithm is done as follows.
As long as there exists a node $v \in V(G')$ such that $|B(v, k d, G')| < d \cdot \ell^{1-1/k}$, pick $v$ to be such a node.
Let $i$ be the minimal index such that $|B(v, i  d , G')| < d \cdot \ell^{(i-1)/k}$.
Construct a shortest-path tree $T(v)$ rooted at $v$ and spanning $B(v, i  d , G')$ and add the edges of $T(v)$ to $H_{d}$.
If $i> 1$, then remove the nodes $B(v, (i-1) d, G')$ from $G'$; if $i = 1$ remove the nodes $B(v, d, G')$.

The second stage of the algorithm is done as follows.
As long as there is a node $v \in V$ such that $B(v,2 k \cdot d,G') \cap C = \emptyset$, pick $v$ to be such a node, and add it to $C$.
For every node $c \in C$, do the following.
First, let $B(c)$ be all nodes $u$ in $G'$ such that $c$ is closer to $u$ than any other $c' \in C$.
(We assume unique shortest paths. This is without loss of generality, as one can artificially create differences between the paths by slightly perturbing the input to ensure uniqueness.)
Second, construct a BFS tree rooted at $c$ and spanning the nodes in $B(c)$ and add the edges of the BFS to $H_d$.
Third, for every label $\lambda \in L$, if there exists a node $v \in B(c)$ such that $\dist(v,\lambda,G) \leq d$, then pick $v$ to be such a node and add a shortest path $P(v,\lambda)$ from $v$ to its closest $\lambda$-labeled node, and add the edges of the path $P(v,\lambda)$ to $H_d$.

This completes the construction of the spanner.
See Figure \ref{VL-Spanner-Unweighted} for the formal code.


\begin{figure}[ht]
\begin{center}
\framebox{\hspace{0.6cm}\parbox{5.5in}{
{\tt Procedure}$\;$ \VLCOVER$(G,d)$
\\
\\
$G' \gets G$, $H_d \gets(V,\emptyset)$, $C \gets \emptyset$
\\
\\
**** Stage 1 ****
\\
while $\exists$ node $v \in V(G')$ such that $|B(v, k d, G')| < d \cdot \ell^{1-1/k}$ do:
\\ $\null\quad$
let $i$ be the minimal index such that $|B(v, i d, G')| < d \cdot \ell^{(i-1)/k}$
\\ $\null\quad$
construct a shortest-path tree $T(v)$ rooted at $v$ and spanning $B(v, i d, G')$
\\ $\null\quad$
add the edges of $T(v)$ to $H_{d}$
\\ $\null\quad$
if $i > 1$ then
\\ $\null\quad\quad$
remove the nodes $B(v, (i-1) d, G')$ from $G'$
\\ $\null\quad$
else (i = 1)
\\ $\null\quad\quad$
remove the nodes $B(v, d, G')$ from $G'$
\\
\\
**** Stage 2 ****
\\
while $\exists v \in V(G')$ such that $B(v,2 k \cdot d,G') \cap C = \emptyset$ do:
\\ $\null\quad$
$C \gets C \cup\{v\}$
\\
for every node $c \in C$ do:
\\ $\null\quad$
let $B(c) = \{u \in V(G') \mid \dist(u,c) = \dist(u,C)\}$
\\ $\null\quad$
construct a shortest-path tree $T(c)$ rooted at $c$ and spanning $B(c)$
\\ $\null\quad$
add the edges of $T(c)$ to $H_d$
\\ $\null\quad$
for every label $\lambda \in L$ such that $\exists y \in B(c)$ such that $\dist(y,\lambda) \leq d$ do:
\\ $\null\quad\quad$
pick $y$ to be such a node
\\ $\null\quad\quad$
add $E(P(y,\lambda,G))$ to $H_d$
\\
return $H_d$
}\hspace{0.6cm}}
\end{center}
\caption{\label{VL-Spanner-Unweighted} Constructing vertex-labeled spanners for unweighted graphs}
\end{figure}

We now turn to analyze the stretch and the number of edges in the resulting spanner $H_d$.


Consider a node $v$ that is picked in the ``while'' loop of the first stage of the algorithm.
Let $G'(v)$ be the graph $G'$ in the algorithm just before $v$ (and the ball around it) was removed from $G'$.

\begin{lemma}
\label{lem:spanner-size}
The number of edges in $H_d$ is $O(n \ell^{1/k})$.
\end{lemma}
\Proof
The algorithm adds edges in three different locations.

The first location is in the first stage of the algorithm.
The second location is edges in the shortest-path trees spanning $B(c)$ for every $c \in C$.
The third location is edges on paths $P(y,\lambda)$ for some node $c \in C$, $\lambda \in L$, and $y \in B(c)$.
We now show that the number of edges added in each of the three locations is $O(n \ell^{1/k})$.

Consider the first location.
Let $v\in V$ be a node that is picked in the first stage of Algorithm $\VLCOVER$ and let $1 \leq i \leq k$ be the minimal index such that $|B(v,i  d, G'(v))| \leq d \cdot \ell^{(i-1)/k}$.
A BFS tree $T(v)$ rooted at $v$ and spanning $B(v,i  d, G'(v))$ is added to $H_d$.
Namely, $|B(v,i  d, G'(v))| < d \cdot \ell^{(i-1)/k}$ edges are added to $H_d$.
We consider two cases, first, when $i > 1$, and second, when $i=1$.
Consider the first case.
Note that by the minimality of the index $i$, $|B(v,(i-1)  d, G'(v))| > d \cdot \ell^{(i-2)/k}$.
We thus can charge the nodes in $B(v,(i-1) \cdot d,G')$ with the edges in $T(v)$ added to $H_d$.
Note that every node in $B(v,(i-1) \cdot d,G')$ is charged with at most $\ell^{1/k}$ edges.
Moreover, the nodes in $B(v,(i-1) \cdot d,G')$ are removed from $G'$ and thus no node is charged twice.
Consider the second case.
Recall that in this case, a BFS tree $T(v)$ is added to $H_d$.
We can charge the nodes in $B(v, d,G')$ with the edges in $T(v)$ added to $H_d$, charging each node with a single edge.
Moreover, the nodes in $B(v,d,G')$ are removed from $G'$ and thus no node is charged twice.
We thus conclude that the number of edges added in the first location is $O(n \ell^{1/k})$.

Consider the second location.
Note that every node belongs to exactly one set $B(c)$ for some $c \in C$.
In addition, note that due to unique shortest paths, for every node $u \in B(c)$, $V(P(u,c)) \subseteq B(c)$.
We thus add a single edge for every node $v\in V$ for this stage.
We obtain that $O(n)$ edges are added for the second location.

Finally, consider the third location.
Let $c\in C$.
Note that $|B(c,k \cdot d,G')| \geq d \cdot \ell^{(k-1)/k}$.
The number of edges added for a path $P(y,\lambda)$ for some $\lambda \in L$ and $y\in B(c)$ is at most $d$, since $\dist(y,\lambda) \leq d$.
There are at most $\ell$ labels, therefore, at most $\ell \cdot d$ edges are added for the node $c$.
We charge the nodes in $B(c,k \cdot d,G')$ with these edges, charging each node in $B(c,k \cdot d,G')$ with at most $O(\ell^{1/k})$ edges.
Note that since the nodes in $C$ are at distance at least $2 k d +1$ from one another, no node is charged twice.
We thus conclude that the number of edges added for the third location is $O(n \ell^{1/k})$.

The lemma follows.
\QED

\begin{lemma}
\label{lem:spanner-d}
For every node $u \in V$ and label $\lambda \in L$ such that $\dist(u,\lambda) \leq d$,
$\dist(u,\lambda, H_d) \leq (4k+1)d$.
\end{lemma}
\Proof
Consider a node $u \in V$ and a label $\lambda \in L$ such that $\dist(u,\lambda) \leq d$.

Let $P(u,\lambda)$ be the shortest path from $u$ to its closest $\lambda$-labeled node $u_{\lambda}$.
We consider two cases.
First case is when some node $y \in V(P(u,\lambda))$ is deleted from the graph $G'$.
The second case is when none of the nodes on $P(u,\lambda)$ is deleted from the graph $G'$.

Consider the first case.
Let $y$ be the first node on $P(u,\lambda)$ that is deleted from $G'$ by the algorithm.
Let $v$ be the node picked by the ``while'' loop of the first stage of the algorithm, such that $y$ is removed in $v$'s iteration.
Let $i$ be the minimal index such that $|B(v, i  d, G')| < x \cdot \ell^{(i-1)/k}$.

We consider two subcases. First, when $i=1$ and second, when $i>1$.
Consider the first subcase.
Note that $V(P) \subseteq V(G'(v))$, since $y$ is the first node in $V(P)$ that is removed from $G'$.
We claim that $V(P) \subseteq B(v,d, G'(v))$, and show it as follows.
Assume, for the sake of argument, that $V(P) \nsubseteq B(v,d, G'(v))$.
This implies that there there must exist a node $w$ such that $\dist(v,w, G'(v)) =d$.
But, note that the shortest path $\tilde{P} = P(v,w, G'(v))$ from $v$ to $w$ in $G'(v)$ contains $d$ nodes and that $V(\tilde{P}) \subseteq B(v,d, G'(v))$.
We thus get that $|B(v,d, G'(v))| \geq d$, contradiction.
Recall that a BFS tree $T(v)$ rooted at $v$ and spanning $B(v,d,G')$ is added to $H_d$.
We get that $\dist(u,\lambda, H_{d}) \leq \dist(u,\lambda, T(v)) \leq
\dist(u,v, T(v)) + \dist(v,u_{\lambda}, T(v)) \leq 2 d$.

Consider the second subcase.
Note that before $B(v,(i-1) \cdot d,G')$ is removed from $G'$, $V(P(u,\lambda)) \subseteq V(G'(v))$, since we assume that $y$ is the first node on $P(u,\lambda)$ that is removed from $G'$.
Recall that, a BFS tree $T(v)$ rooted at $v$ and spanning $B(v,i \cdot d,G')$ is added to $H_d$.
Note also that $V(P(u,\lambda)) \subseteq B(v,i \cdot d,G')$ since $y \in B(v,(i-1) \cdot d,G')$ and $\dist(y,z) \leq d$ for every $z \in V(P(u,\lambda))$.
Specifically, $u,u_{\lambda} \in B(v,i d, G'(v))$.
We get that, $\dist(u,\lambda, H_d) \leq \dist(u,\lambda, T(v)) \leq \dist(u,v, T(v)) + \dist(v,{u_\lambda}, T(v))  \leq 2 i \cdot d \leq 2k \cdot d$.

Consider the second case. We consider again two subcases.
The first subcase is when $u \in C$.
The second subcase is when $u \notin C$.

Consider the first subcase.
Note that $\dist(u,\lambda) \leq d$ and clearly $u \in B(u)$.
Therefore, a path $P(y,\lambda)$ is added to $H_{d}$ for some $y \in  B(u)$.
We get that, $\dist(u,\lambda, H_{d}) \leq \dist(u,y, H_{d}) +\dist(y,\lambda, H_d) \leq 2kd + d = (2k+1)d$.

Consider the second subcase.
The node $u$ is not removed from $G'$, and furthermore $u \notin C$.
Note that this could only happen when $u$ is at distance at most $2 k d$ from some node in $C$.
Let $c \in C$ be the closest node to $u$.
Note that $u \in B(c)$ and thus a shortest path from $u$ to $c$ is added to $H_d$ (as part of $T(c)$).
Moreover, $\dist(u,\lambda) \leq d$.
Hence, a shortest path  $P(y,\lambda)$ is added to $H_d$ for some $y \in B(c)$.
We get that, $\dist(u,\lambda, H_d) \leq \dist(u,c, H_d) + \dist(c,y, H_d) + \dist(y,\lambda, H_d) \leq 2kd+2kd+d = (4k+1)d$.
\QED

The main algorithm for constructing our spanner operates in $\log{n}$ iterations.
For a given fixed parameter $\epsilon$, for every index $1 \leq i \leq \log{n}$, invoke Algorithm \VLCOVER~ with parameter $d(i) = (1+\epsilon)^i$. Let $H_{d(i)}$ be the subgraph returned by the Algorithm \VLCOVER.
Let $H$ be the union of all subgraphs $H_{d(i)}$ for  $1 \leq i \leq \log{n}$.
This completes our spanner construction.
\par
It is not hard to verify that by Lemmas \ref{lem:spanner-size} and \ref{lem:spanner-d}, we have the following.

\begin{theorem}
\label{the:unweight-spanner}
For every unweighted graph $G$ and fixed parameter $\epsilon$, one can efficiently construct a vertex-label $(4k+1)(1+\epsilon)$-spanner
with $O(\log{n} \cdot n \ell^{1/k})$ edges.
\end{theorem}

\subsection{Weighted Graphs}
\label{sec:w-spanners}
In this section we generalize our spanner construction for weighted graphs.

Note that in the unweighted case we exploit the fact that a path $P$ of length $d$ contains $d$ edges.
In the weighted case, this is no longer the case. A path of length $d$ could potentially contain a much smaller or larger number of edges.
We thus need to be more careful with the paths we add to the spanner.
Roughly speaking, for every potential distance $d$ and index $j$, we consider nodes that have at least $2^j$ nodes at distance $d$ from them.
In this case, we can tolerate adding paths with $O(2^j)$ number of edges.


For a path $P$, let $|P|$ be the number of edges in $P$ and let $\dist(P)$ be the length of $P$.
Let $\tilde{\dist}(v,u,x',H)$ be the minimal length of a path from $u$ to $v$ in $H$ among all paths with at most $x'$ edges.
Let $\tilde{P}(u,v,x',H)$ be the shortest path in $H$ between $u$ and $v$ among all paths with at most $x'$ edges.
We say that a node $v$ is $(x',d)$-relevant in $H$ if $x' \leq |B(v,d,H)|$.
We say that a path $P$ is $(x',d)$-relevant if $x' \leq |E(P)| \leq 2x'$ and $\dist(P) \leq d$.

As in the unweighted case, we first describe an algorithm named \WVLCOVER~ that given a distance $d$, an integer $x$, and a graph $G$, returns a subgraph $H_{d,x}$ that satisfy the following.
For every node $v$ and label $\lambda \in L$ such that there exists an $(x,d)$-relevant path $P$ from $v$ to a $\lambda$-labeled node, $\dist(v,\lambda, H_{d,x}) \leq (4k+1)d$.

The algorithm proceeds as follows.
Initially, set $G' \gets G$, $H_{d,x} \gets(V,\emptyset)$, and $C \gets \emptyset$.
There are two stages.
The first stage of the algorithm is as follows.
As long as there exists an $(x,d)$-relevant node $v$ in $G'$ such that $|B(v, k d, G')| < x \cdot \ell^{1-1/k}$,
pick $v$ to be such a node.
Let $i$ be the minimal index such that $|B(v, i  d, G')| < x \cdot \ell^{(i-1)/k}$.
Construct a shortest-path tree $T(v)$ rooted at $v$ and spanning $B(v, i  d, G')$, and then add the edges of $T(v)$ to $H_{d,x}$.
Finally, remove the nodes $B(v, (i-1) d, G')$ from $G'$.

The second stage of the algorithm is done as follows.
As long as there exists an $(x,d)$-relevant node $v$ in $G'$ such that $B(v,2 k \cdot d , G') \cap C = \emptyset$, add $v$ to $C$.
For every node $c \in C$ do the following. First, let $B(c) = \{u \in V(G') \mid \dist(u,c) = \dist(u,C)\}$
(recall that we assume unique shortest paths).
Second, construct a shortest-path tree $T(c)$ rooted at $c$ and spanning $B(c)$, and then add the edges of $T(c)$ to $H_{d,x}$.
Finally, for every label $\lambda \in L$ such that $\exists y \in B(c)$ and $\tilde{\dist}(y,\lambda, 2x) \leq d$,
add $E(\tilde{P}(y,\lambda,2x,G))$ to $H_{d,x}$.
This completes the construction of our spanner.
See Figure \ref{VL-Spanner-Weighted} for the formal code.


%


\begin{figure}[ht]
\begin{center}
\framebox{\hspace{0.6cm}\parbox{5.5in}{
{\tt Procedure}$\;$ \WVLCOVER$(G,d,x)$
\\
\\
$G' \gets G$, $H_{d,x} \gets(V,\emptyset)$, $C \gets \emptyset$
\\
\\
**** Stage 1 ****
\\
while $\exists (x,d)$-relevant node $v \in G'$ such that $|B(v, k d, G')| < x \cdot \ell^{1-1/k}$ do:
\\ $\null\quad$
let $i$ be the minimal index such that $|B(v, i  d, G')| < x \cdot \ell^{(i-1)/k}$
\\ $\null\quad$
construct a shortest-path tree $T(v)$ rooted at $v$ and spanning $B(v, i d, G')$
\\ $\null\quad$
add the edges of $T(v)$ to $H_{d,x}$
\\ $\null\quad$
remove the nodes $B(v, (i-1) d, G')$ from $G'$
\\
\\
**** Stage 2 ****
\\
while $\exists$ $(x,d)$-relevant $v \in V(G')$ such that $B(v,2 k \cdot d, G') \cap C = \emptyset$ do:
\\ $\null\quad$
$C \gets C \cup \{v\}$
\\
for every node $c \in C$ do:
\\ $\null\quad$
let $B(c) = \{u \in V(G') \mid \dist(u,c) = \dist(u,C)\}$
\\ $\null\quad$
construct a shortest-path tree $T(c)$ rooted at $c$ and spanning $B(c)$
\\ $\null\quad$
add the edges of $T(c)$ to $H_{d,x}$
\\ $\null\quad$
for every label $\lambda \in L$ such that $\exists y \in B(c)$ such that $\tilde{\dist}(y,\lambda, 2x) \leq d$ do:
\\ $\null\quad\quad$
pick $y$ to be such a node
 \\ $\null\quad\quad$
add $E(\tilde{P}(y,\lambda,2x,G))$ to $H_{d,x}$
\\
return $H_{d,x}$
}\hspace{0.6cm}}
\end{center}
\caption{\label{VL-Spanner-Weighted} Constructing vertex-labeled spanners for weighted graphs}
\end{figure}


\begin{lemma}
\label{lem:w-spanner-d}
For every node $u \in V$ and label $\lambda \in L$ such that there exists an $(x,d)$-relevant path $P$ from $u$ to a $\lambda$-labeled node,
$\dist(u,\lambda, H_{d,x}) \leq (4k+1)d$.
\end{lemma}
\Proof

Consider a node $u \in V$ and a label $\lambda \in L$ such that there exists an $(x,d)$-relevant path $P$ from $u$ to a $\lambda$-labeled node $u_{\lambda}$.


We consider two cases.
The first case is when some node $y \in V(P)$ is removed from $G'$ in the first stage of the algorithm.
The second case is when none of the nodes in $V(P)$ is removed from $G'$ in the first stage of the algorithm.

Consider the first case.
Let $y$ be the first node in $V(P)$ that is removed from $G'$ by the algorithm.
Let $v$ be the node that is picked by the ``while'' loop of the first stage of the algorithm such that
$B(v,(i-1) d, G'(v))$ is removed from $G'$ and that $y \in B(v,(i-1) d, G'(v))$.
Note that $V(P) \subseteq V(G'(v))$, since $y$ is the first node in $V(P)$ that is removed from $G'$.
Recall that a shortest-path tree $T(v)$, rooted at $v$ and spanning $B(v,i d, G'(v))$, is added to $H_{d,x}$.
Note also that $V(P) \subseteq B(v,i d, G'(v))$.
To see this, recall that $y \in B(v,(i-1) d, G'(v))$, namely, $\dist(v,y,G'(v)) \leq (i-1)d$.
In addition, $\dist(y,z,G'(v)) \leq d$ for every $z \in V(P)$.
Hence, $\dist(v,z,G'(v)) \leq id$ for every $z \in V(P)$.
Specifically, $u,u_{\lambda} \in B(v,i d, G'(v))$.
We get that $\dist(u,\lambda, H_{d,x}) \leq \dist(u,\lambda, T(v)) \leq
\dist(u,v, T(v)) + \dist(v,u_{\lambda}, T(v)) \leq 2i \cdot d \leq 2k \cdot d$, as required.

Let $G''$ be the graph $G'$ at the end of the first phase.

Consider the second case.
We consider two subcases.
First when $u \in C$, and second, when $u \notin C$.

Consider the first subcase.
Note that $\tilde{\dist}(u,\lambda,2x) \leq d$ and clearly $u \in B(u)$.
Therefore, a path $\tilde{P} = \tilde{P}(y,\lambda,2x,G)$ is added to $H_{d,x}$ for some $y \in  B(u)$.
Thus, $\dist(u,\lambda, H_{d,x}) \leq \dist(u,y, H_{d,x}) +\dist(y,\lambda, H_{d,x}) \leq 2kd + d = (2k+1)d$.

Consider the second subcase, where the node $u \notin C$.
Note that $V(P) \subseteq V(G'')$, since none of the nodes in $V(P)$ is removed from $G'$ by the algorithm.
In addition, $V(P) \subseteq  B(u, d, G'')$.
Therefore, $|B(u, d, G'')| \geq |V(P)| \geq x$.
Hence, by definition, $v$ is $(x,d)$-relevant in $G''$.
The node $u$ is not added to $C$ by the algorithm, this could happen only if there is another node $v \in C$
such that $\dist(v,u,G')  \leq 2k\cdot d$.
Let $c \in V(G'')$ be the node such that $u \in B(c)$.
Note that $\dist(c,u,G) \leq \dist(c,u,G'')  \leq 2k \cdot d$.
Moreover, $\tilde{\dist}(u,\lambda, 2 x,G) \leq d$.
Therefore, a path $\tilde{P}(y,\lambda,2x,G)$ is added to $H_{d,x}$ for some $y \in  B(c)$.
We get that, $\dist(u,\lambda, H_{d,x}) \leq \dist(u,c, H_{d,x}) + \dist(c,y, H_{d,x})+\dist(y,\lambda, H_{d,x}) \leq 2kd + 2kd + d = (4k+1)d$.
\QED

\begin{lemma}
\label{lem:w-spanner-s}
The number of edges in $H_{d,x}$ is $O(n \ell^{1/k})$.
\end{lemma}
\Proof
The algorithm adds edges in three different locations.
The first location is in the first stage of the algorithm.
The second location is edges in the shortest-path trees spanning $B(c)$ for every $c \in C$.
The third location is edges on paths $\tilde{P}(y,\lambda,2x,G)$ for some node $c \in C$, $\lambda \in L$, and $y \in B(c)$.
We now show that the number of edges added in each of the three locations is $O(n \ell^{1/k})$.

Consider the first location.
Let $v\in V$ be a node that is picked in the first stage of Algorithm $\WVLCOVER$ and let $1 \leq i \leq k$ be the minimal index
such that $|B(v,i  d, G'(v))| \leq d \cdot \ell^{(i-1)/k}$.
A shortest-path tree $T(v)$ rooted at $v$ and spanning $B(v,i  d, G'(v))$ is added to $H_{d,x}$.
Namely, $|B(v,i  d, G'(v))| < O(x \cdot \ell^{(i-1)/k})$ edges are added to $H_{d,x}$.
Notice that $i>1$. This is due to the fact that $v$ is $(x,d)$-relevant in $G'(v)$.
By the minimality of the index $i$, we have $|B(v,(i-1) \cdot d, G'(v))| > x \cdot \ell^{(i-2)/k}$.
We thus can charge the nodes in $B(v,(i-1) \cdot d, G'(v))$ with the edges in $T(v)$ added to $H_{d,x}$.
Note that every node in $B(v,(i-1) \cdot d, G'(v))$ is charged with at most $O(\ell^{1/k})$ edges.
Moreover, the nodes in $B(v,(i-1) \cdot d, G'(v))$ are removed from $G'$, and thus no node is charged twice.
We thus conclude that the number of edges added for this type is $O(n \ell^{1/k})$.

Consider the second location.
Note that every node belongs to exactly one set $B(c)$ for some $c \in C$.
We thus add a single edge for every node $v\in V$ for this stage.
We obtain that $O(n)$ edges are added for the second location.

Finally, consider the third location.
Let $v\in C$.
Since $v$ is $(x,d)$-relevant in $G''$ and since $v$ is not removed in the first stage of the algorithm, $|B(v, k \cdot d, G'')| > x \cdot \ell^{(k-1)/k}$.
The number of edges added for a path $\tilde{P}(y,\lambda,2x,G)$ for some $\lambda \in L$ and $y \in B(c)$ is at most $2\cdot x$.
Since there are at most $\ell$ labels, at most $2 \ell \cdot x$ edges are added for the node $v$.
We charge the nodes in $B(v, k \cdot d, G'')$ with these edges, charging each node in $B(v, k \cdot d, G'')$ with at most $O(\ell^{1/k})$ edges.
Note that since the algorithm adds the node $v$ to $C$, no node $u$ at distance less than $k \cdot d$ from some node $y' \in B(v, k \cdot d, G'')$ will be picked by the algorithm at some later step.
Thus every node is charged only once.
We can thus conclude that the number of edges added for the third location is $O(n \ell^{1/k})$.

The lemma follows.
\QED

The main algorithm for constructing our spanner operates in $\log{n} \cdot \log{D}$ iterations, where $D$ is the diameter of the graph.
For a given fixed parameter $\epsilon$, for every index $1 \leq i \leq \log{D}$ and $1 \leq j \leq \log{n}$, invoke Algorithm
$\WVLCOVER$ with parameters $d(i) = (1+\epsilon)^i$ and $x(j) = 2^j$. Let $H_{d(i),x(j)}$ be the subgraph returned by the Algorithm $\WVLCOVER$.
Finally, let $H$ be the union of all subgraphs $H_{d(i),x(j)}$ for $1 \leq i \leq \log{D}$ and $1 \leq j \leq \log{n}$.
This completes our spanner construction.

The following lemma shows that the stretch of the spanner is $(4k+1)(1+\epsilon)$.

\begin{lemma}
\label{lem:w-spanner-stretch}
For every node $u \in V$ and label $\lambda \in L$,  $\dist(u,\lambda, H) \leq (4k+1)(1+\epsilon)\dist(u,\lambda, G)$.
\end{lemma}
\Proof
Consider a node $u \in V$ and a label $\lambda \in L$.
Let $P = P(u,\lambda)$ be the shortest path from $u$ to $\lambda$ in $G$.
Let $i$ and $j$ be the indices such that $(1+\epsilon)^{i-1} \leq \dist(P) \leq (1+\epsilon)^i$ and $2^{j} \leq |P| \leq 2^{j+1}$.
By Lemma \ref{lem:w-spanner-d}, $\dist(u,\lambda,H_{d,x}) \leq (4k+1)(1+\epsilon)^i$, for $d = (1+\epsilon)^i$ and $x = 2^{j}$.
The lemma follows.
\QED

We thus conclude Theorem \ref{the:weight-spanner}.
%
%



\end{document}